# Identify main-sequence binaries from the Chinese Space Station Telescope Survey with machine learning

Jia-jia Li,[1,2,5]★ Jin-liang Wang,[1,2] Kai-fan Ji,[1,2] Chao Liu[2,3], Hai-liang Chen,[1,2] Zhan-wen Han[1,2,4] and Xue-fei Chen[1,2,4,5]★

[1]*Yunnan Observatories, Chinese Academy of Sciences, Kunming 650011, P. R. China*
[2]*School of Astronomy and Space Science, University of Chinese Academy of Sciences, Beijing 100049, P. R. China*
[3]*Key Laboratory of Space Astronomy and Technology, National Astronomical Observatories, Chinese Academy of Sciences, Beijing 100101, P. R. China*
[4]*Center for Astronomical Mega-Science, Chinese Academy of Sciences, Beijing 100012, P. R. China*
[5]*International Centre of Supernovae, Yunnan Key Laboratory, Kunming 650216, P. R. China*

**ABSTRACT**
The statistical properties of double main sequence (MS) binaries are very important for binary evolution and binary population synthesis. To obtain these properties, we need to identify these MS binaries. In this paper, we have developed a method to differentiate single MS stars from double MS binaries from the Chinese Space Station Telescope (CSST) Survey with machine learning. This method is reliable and efficient to identify binaries with mass ratios between 0.20 and 0.80, which is independent of the mass ratio distribution. But the number of binaries identified with this method is not a good approximation to the number of binaries in the original sample due to the low detection efficiency of binaries with mass ratios smaller than 0.20 or larger than 0.80. Therefore, we have improved this point by using the detection efficiencies of our method and an empirical mass ratio distribution and then can infer the binary fraction in the sample. Once the CSST data are available, we can identify MS binaries with our trained multi-layer perceptron model and derive the binary fraction of the sample.

.

## 1 INTRODUCTION

Observations show that more than half of the stars are in binary systems (Heintz 1969; Abt & Levy 1976; Duquennoy & Mayor 1991). Binary interaction may explain many astronomical phenomena and form many important objects. For example, type Ia supernovae are generally used as standard candles to measure cosmological distances, and the progenitors of type Ia supernovae are produced by binary evolution, either the single-degenerate model or the double-degenerate model (e.g. Wang & Han 2012). Besides, binary evolution is one of the most important ways to form double stellar-mass black holes, double neutron stars, black hole-neutron star binaries, and double white dwarfs, which are the main sources of gravitational wave radiation detected by both the ground-borne and space-borne gravitational wave detectors, such as LIGO (LIGO Scientific Collaboration et al. 2015), Virgo (Acernese et al. 2015), KAGRA (Kagra Collaboration et al. 2019), LISA (Amaro-Seoane et al. 2022). Binaries are therefore of great importance in modern astrophysics (Han et al. 2020).

Binary population synthesis (BPS) is a universal method for studying the formation of particular stars in the era of big data, and has been widely used to study many objects in the last two decades i.e double white dwarfs (e.g. Han 1998; Nelemans et al. 2001), progenitors of SNe Ia (e.g. Yungelson & Livio 1998; Wang & Han 2012), double NSs (e.g. Bethe & Brown 1998; Portegies Zwart & Yungelson 1998; Bloom, Sigurdsson & Pols 1999; Belczynski, Kalogera & Bulik 2002), binary black holes (e.g. Lipunov, Postnov & Prokhorov 1997; De Donder & Vanbeveren 1998), X-ray binaries (e.g. Pfahl, Rappaport & Podsiadlowski 2003), hot subdwarfs (e.g. Han et al. 2002, 2003), blue stragglers (e.g. Chen & Han 2008), symbiotic stars (e.g. Yungelson et al. 1995; Lü, Yungelson & Han 2006), etc. Binary fractions and distributions of mass ratio, orbital period, eccentricity, etc., are fundamental inputs of the BPS method and would affect the final results significantly. Such statistical properties can also be used to constrain star formation rate, but they have not been well investigated in the last century due to the sample size limit.

In the new Millennium, large sky survey projects include SDSS, APOGEE, LAMOST, and *Gaia*. have boosted the study of statistical properties of binary population (e.g. Clark, Blake & Knapp 2012; Gao et al. 2014; Yuan et al. 2015; Andrews, Chanamé & Agüeros 2017; Badenes et al. 2018; El-Badry & Rix 2019; Mazzola et al. 2020; Hwang, Ting & Zakamska 2022; Li et al. 2022a,b; Guo et al. 2022a, b; Li et al. 2023). However, there is no consensus on such fundamental properties of the binary population yet. For a population, the binary fraction is one of the most important parameters to assess the influence of the binary population. Studies show that $f_b$ varies with effective temperature, metallicity and population age (e.g. Moe & Di Stefano 2017). For example, $f_b$ appears to be proportional to stellar mass (Duchêne & Kraus 2013). It is about 22 per cent (Allen 2007) for very low mass field stars, about 26 per cent (Delfosse et al. 2004) for low-mass stars, about 41–50 per cent (Raghavan et al. 2010) for solar-type stars, and greater than 50 per cent for massive stars (Sana et al. 2013). Some studies show an anticorrelation between binary fraction and metalicity (e.g. Gao et al. 2014), while some studies show a weak positive correlation or no correlation between $f_b$ and metallicity (e.g. El-Badry & Rix 2019).

★ E-mail: lijiajia@ynao.ac.cn (JJL); cxf@ynao.ac.cn (XFC)

The China Space Station Telescope (CSST), scheduled to launch in 2024, will perform one of the most important photometric surveys in the next few years. The CSST will have a 2-m diameter primary mirror and a field of view 1.7 deg$^2$, about 300 times larger than that of the *Hubble Telescope* (Zhan 2011). The primary mirror is designed for photometry and slitless spectroscopy, covering the wavelength range of 255–1000 nm. Seven photometric bands cover from the near-ultraviolet to the near-infrared, i.e. NUV: 2520–3210 Å, *u*: 3210–4010 Å, *g*: 4010–5470 Å, *r*: 5470–6920 Å, *i*: 6920–8420 Å, *z*: 8420–10 800 Å, and *y*: 9270–10 800 Å. The average transmission of visible light and infrared light is greater than 0.65 (Cao et al. 2018). Its spatial resolution is around 0.15 arcsec, and the observation depth can reach 26 mag in the *g* band, 24.4 mag in the *y* band, and 25.5 on average in the remaining bands (Gong et al. 2019).

During the operation of the CSST, it will focus on the the medium and high galactic latitude instead of the galactic disc. It is expected that a lot of globular clusters and field stars but few open clusters will be observed.

The wide field of view of the CSST will allow us to image up to 40 per cent of the sky over ten years, obtaining billions of photometric and spectral data of stars and providing great opportunities for studying binary populations.

The majority of binaries are optically indistinguishable. Photometry observations are usually used to find eclipsing binaries, but they need multiple observations and are orbital inclination dependent. Colour–colour diagram (CCD) and colour–magnitude diagram (CMD) have been used to discover binaries in clusters (e.g. El-Badry et al. 2018; Li et al. 2020; Price-Whelan et al. 2020). Binary stars are mainly located in the brighter and redder directions of the main sequence (MS) in a CMD. Suppose the two binary components have comparable masses, i.e. a mass ratio close to 1.0, the binary appears 0.753 mag redder in the vertical direction of the MS belt (Hurley & Tout 1998). Such binary sequences (BS) have been seen in the CMDs of many clusters (e.g. Montgomery, Marschall & Janes 1993; Geller, Latham & Mathieu 2015; Reino et al. 2018), and some statistical analysis have been done based on this (e.g. Sollima et al. 2007, 2010; Niu, Wang & Fu 2020). The CMD has been divided into binary and single-star regions in these studies. The binary fraction and space distribution would be obtained by counting the stars in each region and tracing their places in the clusters. The mass ratio distribution can also be obtained by comparing the observed CMDs with the synthetic CMDs generated by a population with a given binary fraction and mass ratio distribution (Sollima et al. 2010; Clem et al. 2011; Milone et al. 2012; Li, de Grijs & Deng 2014).

The limit of the CCD and CMD method is that the positions of binary stars usually overlap with that of evolved single stars on the two type diagrams. It means that many stars cannot be confirmed as binaries or single stars only based on the CCDs or CMDs of clusters without spectroscopic observations of each star.

A better way to distinguish binaries from single stars is to use the observations in the seven different filters of the CSST simultaneously. However, there is no obvious correlation between binaries and the seven magnitudes. This paper aims to develop a model to distinguish binaries from single stars from the CSST photometry survey based on a neural network and a deep learning network is constructed based on multi-layer perceptron (MLP). The paper is structured as follows. In Section 2, we introduce how we construct the mock data for our simulation. In Section 3, we introduce the MPL model. In Section 4, we present our simulation results. Discussion and conclusion of this work are given in Sections 5 and 6.

## 2 THE MOCK DATA

We first need a mock sample of stars, which is a mixture of single stars and binaries, to start our simulation. Here we only focus on stars located in the MS, the most challenging and concerned part of the Hertzsprung–Russell diagram (HRD). Below we briefly describe how we construct the sample and compute the magnitudes of these systems:

(1) Construct a mock sample of single MS stars: The stellar mass $M$ in this sample ranges from 0.1 to $10\,M_\odot$ with a step of $\Delta \log M/M_\odot = 0.01$. The metallicity is assumed to be $Z = 0.02$. For a star with a given mass $\geq 0.8\,M_\odot$, we can have an evolutionary track from the stellar track library, Modules for Experiments in Stellar Astrophysics Isochrones & Stellar Tracks (MIST; Choi et al. 2016; Dotter 2016), and then divide the MS phase (i.e. from zero-age MS to the turnoff of the MS)[1] into 100 parts equally in stellar radius, i.e. $\Delta \log R/R_\odot$. Consequently, we have 101 points on the MS phase for each track and the stellar properties at each point, including effective temperature $T_{\rm eff}$, surface gravity $\log g$, and stellar age $t$ can be obtained from the stellar evolution tracks. For stars with masses lower than $0.8\,M_\odot$, the stars have almost not evolved within the Universe age. Therefore, we adopt the stellar properties at the zero-age MS for these stars. In this way, we can get a grid of single stars covering the above mass range and the whole MS phase, shown in Fig. 1(a). We randomly select the stars from this grid to get a mock sample of single stars.

(2) *Construct a mock sample of binaries*. We obtain the sample of the primary as the single star sample. Regarding the secondaries, we can obtain their mass ($M_2$) for a given mass ratio, which is randomly produced from a mass ratio distribution $f(q)$. With the given secondary mass and stellar age,[2] we can obtain its other stellar parameters by interpolating the stellar evolutionary tracks from MIST. In this way, we can obtain a mock sample of binaries, which is shown in panel (b) of Fig. 1.

(3) *Spectra library*. The BT-Settle spectra library (Allard et al. 2013) is used to obtain the spectra of a star with a given effective temperature and surface gravity. The range of effective temperature and surface gravity in the BT-Settle spectra library is $2600 \leq T_{\rm eff} < 50\,000$ K, $0.5 < \log(g/({\rm cm\,s^{-2}})) < 6$ at solar metallicity. In Fig. 1(c), we show our mock sample and the grid in the spectra library. From this plot, we see that the grid of the spectra library can well cover the parameter space of our mock sample.

---

[1] In our simulation, we chose the point when the central H abundance is smaller than 0.1 (0.028) as the turn-off points of MS for stars with masses smaller (larger) than $\sim 1.25\,M_\odot$ by checking the evolutionary tracks.

[2] Here we assume that the primary and secondary stars have the same age.

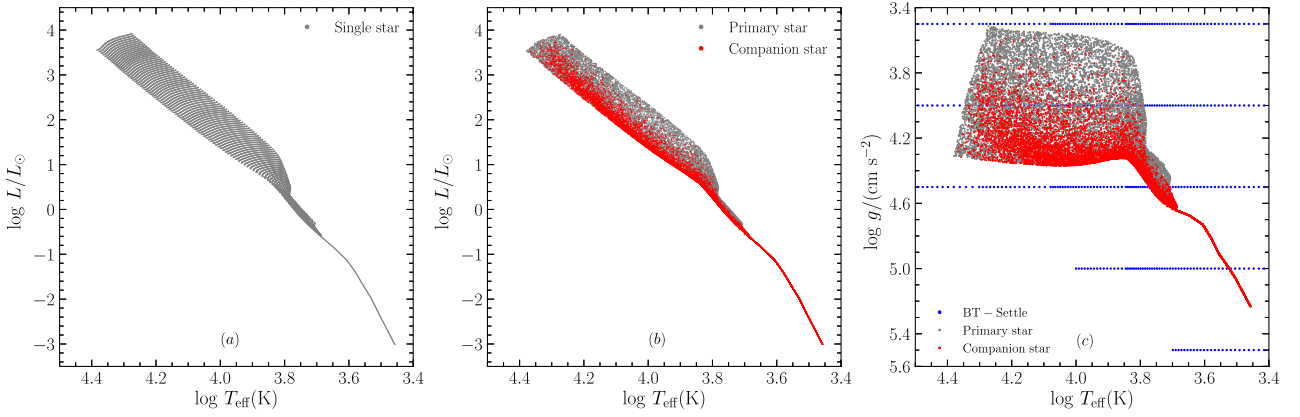

**Figure 1.** (a) Distribution of single star models on the HR diagram. (b) HR diagram of the binary stars in our mock sample. (c) The log $T_{\rm eff}$–log $g$ parameter space for the binary sample (grey and red dots) and the grid in BT-Settle spectra library (blue dots). The grey and red dots in (b) and (c) represent the primary and companion stars, respectively.

(4) *The CSST magnitudes.* We denote the transmission curve of the CSST photometry as $S_{\lambda,i}$, where $i = 1, 2, 3, 4, 5, 6$, and 7 are for the NUV, $u$, $g$, $r$, $i$, $z$, and $y$ bands, respectively.

The absolute magnitude of an object at each band is obtained by

$$M_i = -2.5 \log \left[ \left(\frac{R}{10\text{pc}}\right)^2 \frac{\int_{\lambda_1}^{\lambda_2} \lambda f_\lambda S_{\lambda,i} d\lambda}{\int_{\lambda_1}^{\lambda_2} \lambda f_\lambda^0 S_{\lambda,i} d\lambda} \right]$$

$$= -2.5 \log \left(\frac{R}{10\text{pc}}\right)^2 + m_i, \quad (1)$$

$$m_i = -2.5 \log \left[ \frac{\int_{\lambda_1}^{\lambda_2} \lambda f_\lambda S_{\lambda,i} d\lambda}{\int_{\lambda_1}^{\lambda_2} \lambda f_\lambda^0 S_{\lambda,i} d\lambda} \right], \quad (2)$$

where $f_\lambda^0$ is the flux of the reference spectra and $f_\lambda^0 = (c/\lambda^2) f_\nu^0$, $f_\nu^0 = 10^{48.60/-2.5}$ erg s$^{-1}$ cm$^{-2}$ Hz$^{-1}$, since the AB magnitude system is adopted on the CSST. Parameters $\lambda_1$ and $\lambda_2$ are the wavelength boundaries of each band; $R$ is the stellar radius and $f_\lambda$ is the intrinsic flux.

For an unresolved binary, the two components are too close to be resolved in the image. That means they look like a single-point source with the light of two parts. Therefore, the magnitude of a binary star is the combination of their two single stars. The absolute magnitude of the binary system can be calculated with the following equation:

$$M_{b,i} = -2.5 \log \left[ \frac{R_1^2 \int_{\lambda_1}^{\lambda_2} \lambda f_{\lambda 1} S_{\lambda,i} d\lambda + R_2^2 \int_{\lambda_1}^{\lambda_2} \lambda f_{\lambda 2} S_{\lambda,i} d\lambda}{(10 \text{ pc})^2 \int_{\lambda_1}^{\lambda_2} \lambda f_\lambda^0 S_{\lambda,i} d\lambda} \right]$$

$$= M_{1,i} - 2.5 \times \log \left(1 + 10^{\frac{M_{1,i} - M_{2,i}}{2.5}}\right), \quad (3)$$

where $f_{\lambda 1}$ and $f_{\lambda 2}$ are the fluxes of the two parts in the binary system; $M_{1,i}$ is the absolute magnitude of the primary star and $M_{2,i}$ is the absolute magnitude of the companion star; $R_1$ and $R_2$ are the radius of the primary and secondary, respectively.

Regarding the magnitude errors of the CSST, we estimate it based on the Poisson error of the flux. The relation between the SNR and the errors of magnitudes can be derived as follows.

The definition of magnitude ($m$) is

$$m = -2.5 \log \left(\frac{F}{F_0}\right), \quad (4)$$

where $F$ is the flux of the specific source and $F_0$ is the flux of the reference source. Then the magnitude error ($\sigma_m$) can be calculated as follows.

$$|\sigma_m| = 1.087 \left|\frac{\sigma_F}{F}\right| \approx \left|\frac{\sigma_F}{F}\right| = \frac{1}{SNR}, \quad (5)$$

where $\sigma_F$ is the error of flux.

In our simulations, we consider five choices of SNR, i.e. 20, 50, 100, 200, and $+\infty$, corresponding to errors of magnitude, 0.05, 0.02, 0.01, 0.005, and 0. We assume the errors at different bands are the same and follow a Gaussian distribution.

## 3 THE METHOD

### 3.1 MLP network

MLP is a forward-structured artificial neural network (MLP) that maps a set of input vectors to a set of output vectors. The MLP can be viewed as a direct graph consisting of multiple layers of nodes, where each node is fully connected to the next layer. Every node is a neuron with a non-linear activation function except for the input ones. The MLP is trained using the supervised learning method of the backpropagation (BP) algorithm invented by Geoffrey Hinton (Rumelhart, Hinton & Williams 1986). The BP algorithm uses the Sigmoid function for non-linear mapping, where the problem of non-linear classification and learning is effectively solved.

Fig. 2 shows the frame of our neural network, which has three hidden layers. First of all, each of the $M_i - (\sum_{i=1}^{7} M_i/7)$ (difference between the seven-band magnitude and the seven-band average magnitude[3]) is allocated to a node of the input layer so that a neuron will receive all colour profile information delivered by the notes. The output layer is a binary judgment result of whether the star is a binary or a single star; a binary's output is 1, and the output of a single star is 0. All nodes are fully connected to neurons in the next hidden layer. During the training process, the weights and biases of each neuron are calculated and passed to the subsequent layers after adjustment. The output of each layer is related to the results of previously hidden layers and the initial input. We set up three hidden layers, each placing one hundred neurons. We adopted the classic Adam solver for weight optimization in the network because it works well regarding training time and validation scores on relatively large data sets (containing thousands of training samples or more). We use the ReLU function as the Activation to return the maximum between the input and 0. We use the Cross-Entropy Loss, which is commonly used in machine learning, particularly in classification problems. It measures the difference between the probability distributions of model output and true labels, thus providing a better measure of model accuracy. Other configuration parameters will not be introduced one by one here. We adopt this network because of its good performance after experiments.

---

[3]These kinds of parameters are chosen to avoid the effect of distance. On the other hand, these are equivalent to numerical normalization, which speeds up the gradient descent to find the optimal solution and may also improve the accuracy.

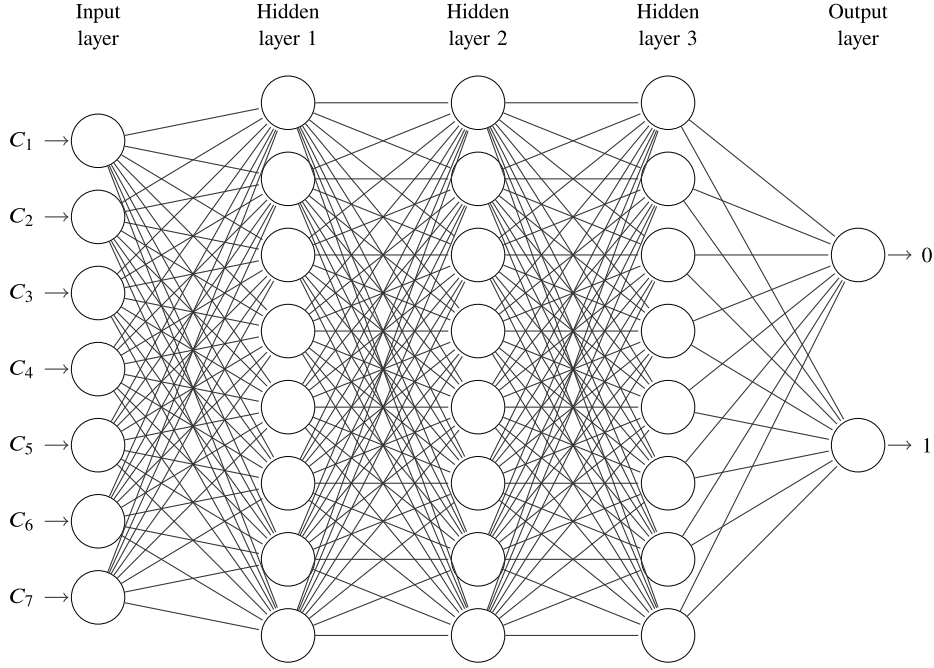

**Figure 2.** The frame of our network. Input layer: $C_i = M_i - (\sum_{i=1}^{7} M_i/7)$, The numbers from 1 to 7 correspond to the seven bands of the CSST, i.e. NUV, $u$, $g$, $r$, $i$, $z$, $y$. Hidden layer: there are three layers in this part and each layer has 100 neurons. Output layer: 1 represents a binary star and 0 represents a single star.

### 3.2 Training the MLP model

We now construct a mock sample following the method described in Section 2 and use it to train our MLP model. The number of systems in the mock sample is 28 572. 70 per cent of them are for the training and the rest are for the test. The training sample has 20 000 systems, consisting of 10 000 single stars and 10 000 binaries. In Section 5.1, we discuss the influence of the number of systems in the mock sample on our results. In addition, we assume that the mass ratio follows a flat distribution for the binary systems. To avoid producing any bias in the mock sample, we also take the single stars as binaries and randomly assign a mass ratio to each single star following the same mass ratio distribution.[4] Given the uncertainties of SNR, we have five mock samples, i.e. for SNR = $+\infty$, 200, 100, 50, 20.

After training, we use the $F_1$ score defined as follows to assess the effect of the training (Chinchor 1992):

$$F_1 = \frac{2 \times P \times R}{P + R}, \qquad (6)$$

---

[4]This is because when dealing with a batch of unknown samples, we are uncertain about which stars are binaries and which are single stars. To address this uncertainty, we assume that all stars are binaries and assign them a mass ratio as one of their properties.

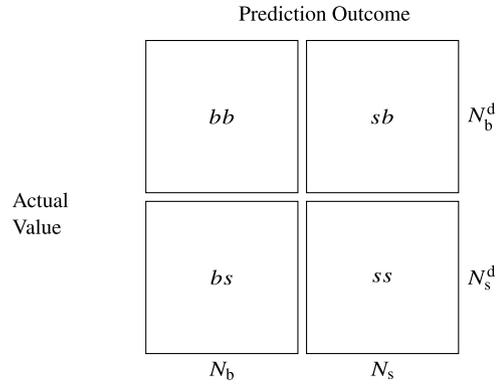

**Figure 3.** Confusion matrix: 'bb' represents the number of binaries, which are detected as binaries in our MLP models; 'sb' represents the number of single stars, which are detected as binaries in our MLP models; and 'bs' represents the number of binaries, which are detected as single stars in our MLP models. 'ss' represents the number of single stars that are detected as single stars in our MLP models.

where $P$ is the fraction of true binaries in all detected binaries, and $R$ is the fraction of the true binaries detected in the total binaries of the mock sample. The $F_1$ score is the harmonic mean of precision and accuracy based on a confusion matrix. The higher the $F_1$ value is, the more reliable the model is.

For each system (i.e. single or binary) in the test sample, we can obtain its probability of being a binary $\rho_b$ (or being a single star $\rho_s$) with the trained MLP model. Those with $\rho_b \geq 0.5$ ($\rho_s \geq 0.5$) will be recognized as a binary (single star) finally. The output can then be divided into four groups as shown in Fig. 3, representing the binaries have been detected as binaries (bb), the single stars have been detected as binaries (sb), the binaries have been detected as single stars (bs), and the single stars have been detected as single stars (ss), respectively. Denoting the number of stars in each group as $N_{bb}$, $N_{sb}$, $N_{bs}$ and $N_{ss}$, we then may write the number of binaries as $N_b = N_{bb} + N_{bs}$, and that of single stars as $N_s = N_{sb} + N_{ss}$ in the mock sample. Then $P = N_{bb}/(N_{bb} + N_{sb})$ and $R = N_{bb}/N_b$.

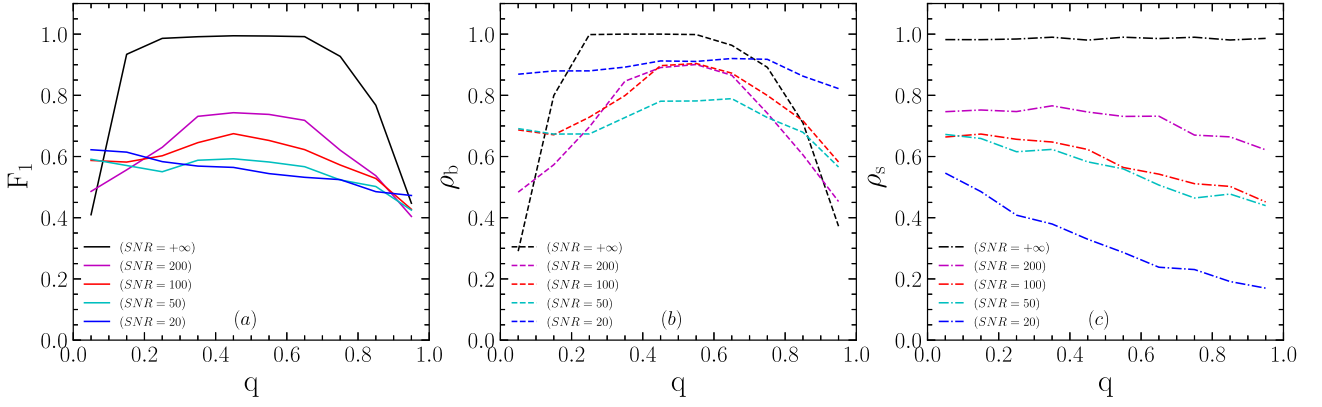

**Figure 4.** Dependence of $F_1$ (a), $\rho_b$ (b), and $\rho_s$ (c) on the binary mass ratio. The SNRs from the upper panels to the lower panels are $+\infty$, 200, 100, 50, and 20, respectively.

For each trained MLP model, we use a binary detection efficiency $\rho_b = N_{bb}/N_b$ and a single star detection efficiency $\rho_s = N_{ss}/N_s$ to assess the ability of the MLP model.

### 3.3 Binary fraction

In a sample with $N$ systems, i.e. $N_b$ binaries and $N_s$ single star ($N = N_b + N_s$), after running our trained MLP models, the number of detected binaries $N_b^d$ can be calculated with the following equations:

$$
\begin{aligned}
N_b^d &= N_{bb} + N_{sb} \\
&= \rho_b N_b + (1-\rho_s)N_s \\
&= \rho_b N_b + (1-\rho_s)(N-N_b) \\
&= (\rho_b + \rho_s - 1)N_b + (1-\rho_s)N.
\end{aligned}
\quad (7)
$$

Then we have the number of binary in the sample $N_b$

$$
N_b = \frac{(1-\rho_s)N - N_b^d}{1 - \rho_b - \rho_s}, \quad (8)
$$

and the binary fraction of the sample $f_b$,

$$
f_b = \frac{1 - \rho_s - f_b^d}{1 - \rho_b - \rho_s}, \quad (9)
$$

where $f_b^d = N_b^d/N$ is the detected binary fraction of the sample.

From equation (8), we can find that $N_b^d$ converges towards $N_b$ for high efficiency, while it may not be a good approximation if the efficiency of the detection decreases. Therefore we need to find a method to infer the value of $N_b$ if the detection efficiency is small.

## 4 RESULTS

### 4.1 Reliability of the MLP models

To study the dependence of $F_1$, $\rho_b$, and $\rho_s$ on $q$ in our MLP model, we divide the test sample into 10 small samples according to mass ratio. For each small sample, we run the trained MLP model and can get the values of $F_1$, $\rho_b$, and $\rho_s$. The dependence of these values on the mass ratio is shown in Fig. 4. For the MLP model with SNR $= +\infty$, we can find that the $F_1$ value is larger ($\sim 0.90$) for binaries with mass ratios $0.2 < q < 0.8$. This means that the trained MLP model is more reliable for these binaries. Moreover, we have $\rho_b \geq 98$ per cent for $0.25 < q < 0.6$, $\rho_b \geq 80$ per cent for $0.2 < q < 0.8$, and $\rho_b \sim 50$ per cent for $q = 0.1$ and $0.9$. When $q$ is close to 0 or 1, $\rho_b$ is only about 25 per cent. This means that binaries with $0.25 < q < 0.60$ would be recognized easily, and it becomes more difficult for binaries with $q$ smaller than 0.25 or larger than 0.60. This could be easily understood as follows. For a small $q$ value, the contribution of flux from the companion is small, while for a large $q$ value, the contribution from the companion becomes similar to that of the primary. Both these factors lead to the binary resembling the primary of a binary if we only consider the relative magnitudes. On the other hand, the $\rho_s$ is independent of the mass ratio, which is expected.

As the decrease of SNR, the value of $F_1$ decreases for these binaries with mass ratio $0.2 < q < 0.8$ and increases for other binaries. Similar trends for $\rho_b$ values have been found. Moreover, the detection efficiency $\rho_s$ decreases as SNR decreases more significantly for systems with larger mass ratios. We can see that the F1 score also becomes smaller at larger mass ratios as the SNR decreases. This may be because there will still be one larger and one smaller star for binary stars with extremely small mass ratios($q \sim 0$). The input amount of machine learning is slightly different, and machine learning can understand such differences well. For extremely large mass ratios($q \sim 1$), the two stars in the binary stars are the same, with no distinction, causing the model to be difficult to classify. It is evident that at the SNR = 200, $\rho_s$ is almost a constant in the $0 < q < 1$. That means the detection efficiency of the model for single stars does not depend on the 'mass ratio of single stars'. This is expected since there is no dependence of the properties of single stars on the mass ratio. However, the reliability of our model decreases with the decrease in SNR.

### 4.2 Influence of mass ratio distribution

In the above simulation, we assume that the mass ratio distributions in the training and test samples are the same, i.e. a uniform distribution. To understand the influence of mass ratio distribution, we fix the trained MLP model but vary the mass ratio distribution in the test sample.

We construct some test samples consisting of 5000 single stars and 5000 binaries. We assume the mass ratio follows a normal, exponential, or negative exponential distribution in these samples, which are shown in Fig. 5. In Fig. 6, we compare results from samples with different mass ratio distributions at SNR = 200. This plot shows the minimal difference between models with different mass ratio distributions. This is because the mass ratio distribution only influences the number of binary systems for a given mass ratio bin but not the physics properties of the binaries. Here we only show the case with SNR = 200 since we find the same conclusion for other cases.

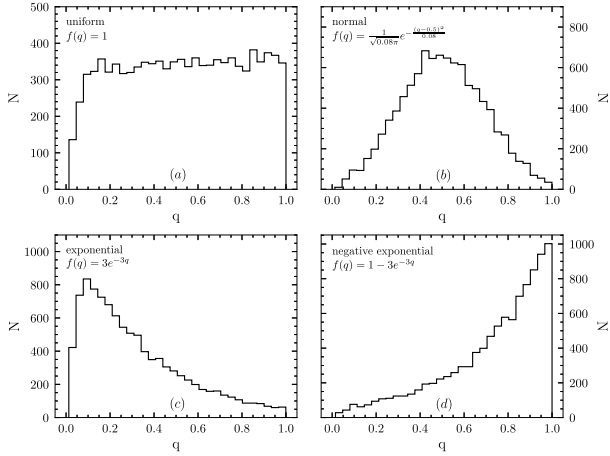

**Figure 5.** Four mass ratio distributions (i.e. uniform, normal, exponential and negative exponential distribution) were adopted in our test sample. In each panel, we also show the corresponding probability density function.

### 4.3 A method to infer the number of binaries and binary fraction with our trained MLP model

If we apply our trained MLP model to a real observational sample, we can get the detected number of binaries ($N_b^d$), which is not always a good approximation to the true number of binaries in the sample (see equation 8). In this section, we develop a method to infer the number of binaries ($N_b$) and binary fraction with our trained MLP model.

We construct several test samples with the number of stars $N$ = 10 000 to represent the observational sample. For each $N$, we vary $N_b$ from 100 to $N - 100$ by a step of 500. And for each ($N$, $N_b$), the test samples are generated with the method described in Section 2. The binary mass ratio is assumed to follow the uniform, exponential, negative exponential and normal distribution.

With these test samples, we run the trained MLP model. We can get the difference between the number of detected binaries ($N_b^d$) and the accurate number of binaries ($N_b$) in the original sample, which is shown in the left column of Fig. 7. From the upper panel, we can find that $N_b^d$ is very close to $N_b$ in the case with a normal mass ratio distribution compared with other cases. This is because there are more binaries with mass ratios $0.2 < q < 0.6$ in the case with a normal mass ratio, and the trained MLP model is very reliable in this mass ratio range (see Fig. 4). With the decrease in SNR, the difference between the $N_b^d$ and $N_b$ becomes larger. This means that it is not bad to use $N_b^d$ to infer the value of $N_b$ in the case with SNR = $+\infty$ but not in other cases. Therefore, we need to find a method to get a more accurate $N_b$ instead of obtaining it from the MLP model.

In reality, we can have prior information about the mass ratio distribution and try to infer the true number of binaries $N_b$ with our trained MLP model in the sample. The basic idea is as follows. We can know the number of binaries in a given mass ratio range for a given mass ratio distribution. Since the detection efficiency is independent of mass ratio distribution (see Fig. 6), we can use the same detection efficiency for different samples. With this detection efficiency and the number of detected binaries within a mass ratio range, we can infer the number of binaries in the mass ratio range from equation (8). Then, we can get the total number binaries in the original sample. In the following, we will use $N_b^{d'}$ to denote the number of binaries inferred with this method.

As the middle column of Fig. 7 shows, in an ideal case, if we know the exact mass ratio distribution of the test sample, we can find that the $N_b^{d'}$ is very close to $N_b$ in different cases following the above idea. As the decrease of SNR, the deviation between $N_b^{d'}$ and $N_b$ becomes larger, particularly for the case with a negative exponential mass ratio distribution. This is mainly because the mass ratio distribution will influence the number of binaries, which are easily found with our trained MLP model. From these plots, we demonstrate that our method should work.

In a more reasonable case, we can not know the exact mass ratio distribution for an observational sample. But we know some empirical mass ratio distributions from previous studies. Following the above idea, we take the uniform mass ratio distribution (i.e. $f(q) = 1.0$) as the empirical distribution to infer the number of binaries for each test sample. The results are shown in the right column of

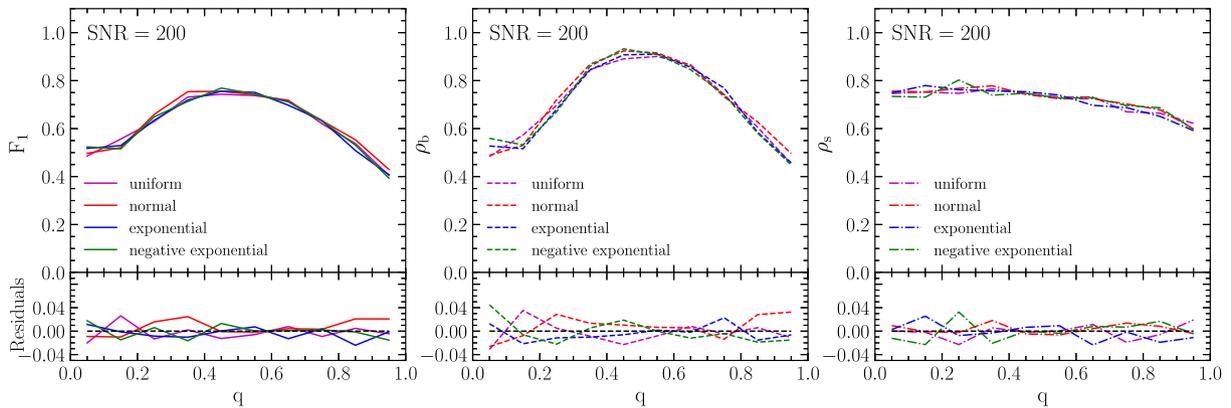

**Figure 6.** Influence of mass ratio distribution on the dependence of $F_1$ (solid line), $\rho_s$ (dotted line), and $\rho_b$ (dashed line) on the mass ratio. The residuals are calculated for each case relative to the average of the four cases with different mass ratio distributions. The purple, red, blue, and green colours are for the cases with uniform, normal, exponential, and negative exponential mass ratio distribution, respectively. Here we only show the case with SNR = 200 since other cases are very similar.

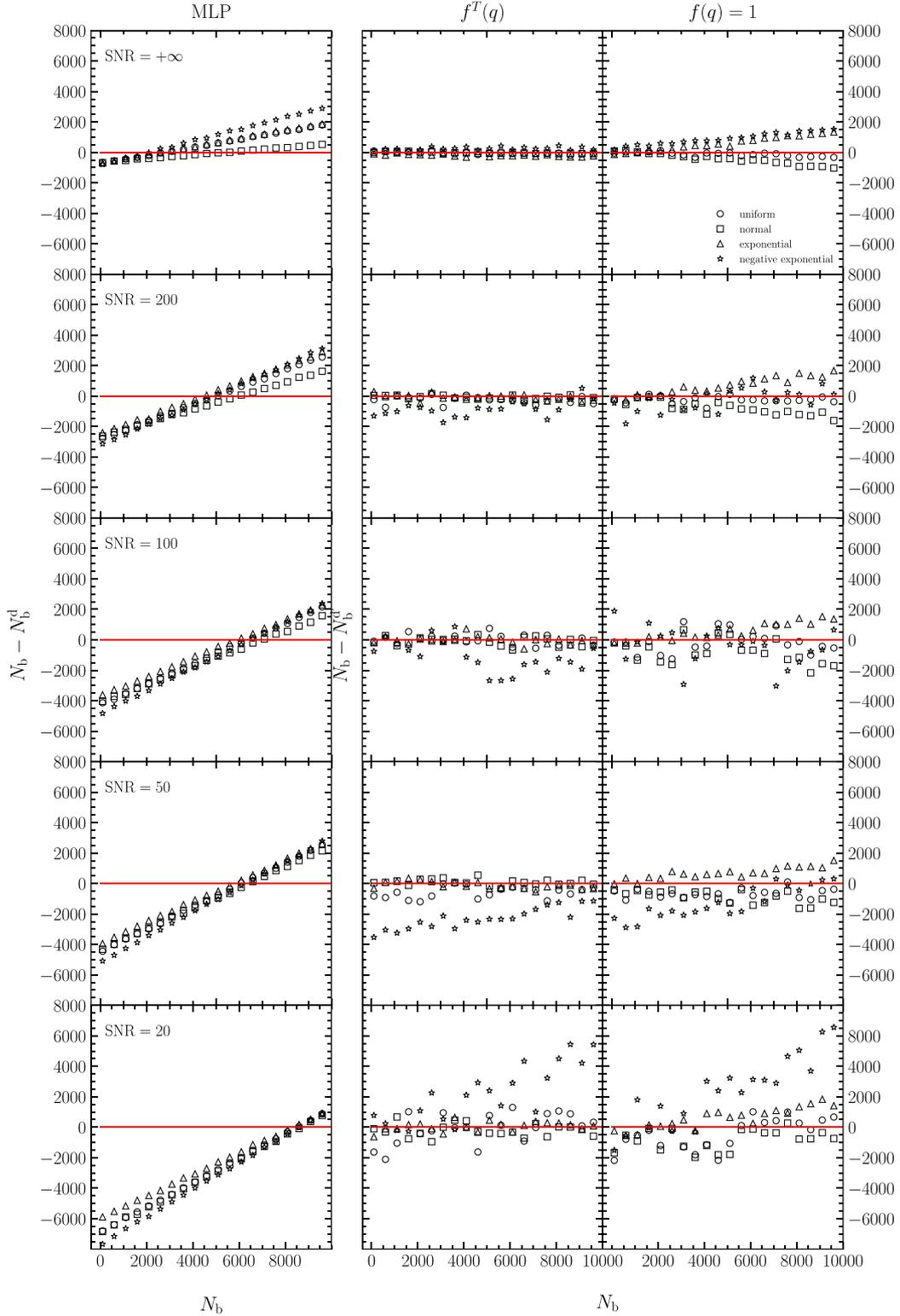

**Figure 7.** Dependence of the difference between the accurate number of binaries and the inferred number of binaries on the accurate number of binaries in a sample. In the left column, the inferred number of binaries is the detected number of binaries from our trained MLP model. In the middle panel, we use the exact the same mass ratio distribution as the test sample to inferred the number of binaries following the method in Section 4.3. In the right column, assuming we have no information of the mass ratio distribution, we take the flat mass ratio distribution to inferred the number of binaries following the method in Section 4.3. The circle, squire, triangle, and star symbols represent uniform, normal, exponential, and negative exponential mass ratio distribution of the test samples, respectively.

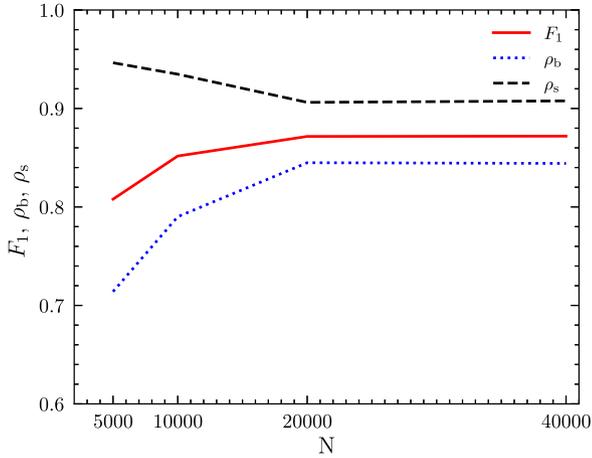

**Figure 8.** Dependence of $F_1$ (red solid line), $\rho_b$ (blue dot line), and $\rho_s$ (black dash line) on the number of systems in the training samples of MLP models.

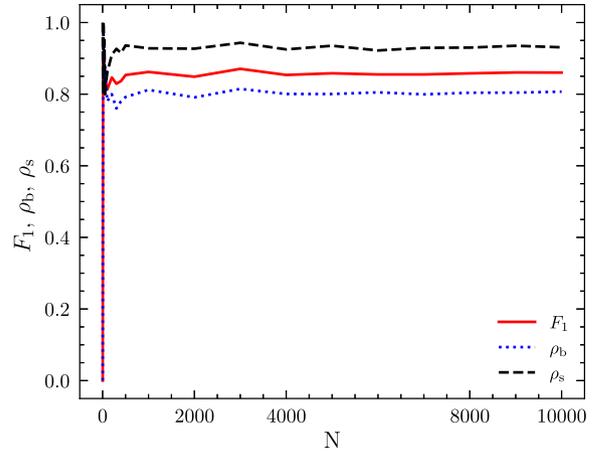

**Figure 9.** Dependence of $F_1$ (red solid line), $\rho_b$ (blue dot line), and $\rho_s$ (black dash line) on the number of systems in the test samples of MLP models.

Fig. 7. From these plots, we can find that generally $N_b^d$ is much closer to $N_b$ in contrast with these plots on the left, even if the mass ratio distributions in the test samples are different from the uniform distribution.

## 5 DISCUSSION

### 5.1 Influence of the sample size

In the above sections, the sizes of the training and test sample do not vary significantly, which is unrealistic. In this part, we study the influence of the sizes of the training and test samples on our results.

We first construct several training samples with different numbers (i.e. $N_{\rm train}$ = 5000, 10 000, 20 000, and 40 000) of systems to train our model and a test sample with 10 000 systems. In these samples, we adopt a balanced mass ratio distribution and a binary fraction of 50 per cent. The dependence of $F_1$, $\rho_b$, and $\rho_s$ is shown in Fig. 8. From this plot, we can find that the values of $F_1$, $\rho_b$, and $\rho_s$ start to converge when $N_{\rm train}$ is larger than 20 000. This explains why we have a sample with 20 000 systems to train our MLP model in the above sections.

To study the influence of the number of systems in the test sample on the values of $F_1$, $\rho_b$, and $\rho_s$, we fix the training sample with 20 000 systems and construct several test samples with different sizes (i.e. $N_{\rm test}$ = 2, 10, 50, 100, 200, 500, 1000, 2000, 3000... 9000, and 10 000). In Fig. 9, we compare the performance of the MLP model from test samples with different sizes. We can find that the values of $F_1$, $\rho_b$, and $\rho_s$ start to converge when the number of systems in the test samples is larger than 500. This also means our method cannot be applied to a small sample.

### 5.2 Application of our method to different environment

It is expected that there will be a lot of globular clusters but few open clusters observed in the CSST survey. The number of stars expected to be detected in a globular cluster is definitely larger than 1000. As shown in Fig. 9, we can apply our method to the stellar population in the galactic field and the globular cluster. With our trained MLP model, we can identify these MS binaries and obtain the binary fraction in different environment. This will be very helpful for us to study the role of dynamical interaction on the formation of binaries.

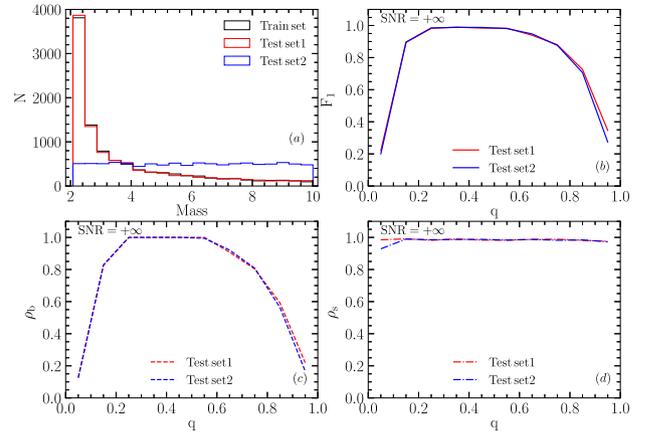

**Figure 10.** (a) The IMFs of single stars and primaries of binaries in the training sample and test samples. In the training (black line) and one test sample (red line), the masses of single stars and primaries of binaries follow a flat distribution in logarithmic space. In the other test sample (blue line), the masses follow a flat distribution. Panels (b), (c), and (d) show the dependence of $F_1$ (solid line), $\rho_b$ (dotted line), and $\rho_s$ (dashed line) on the binary mass ratio for two test samples, respectively.

### 5.3 Influence of the IMF of single MS stars and the primary stars

In our test sample, we assume that the masses of the single stars and the primaries of binaries follow a flat distribution in logarithmic space. However, the IMF of the single stars and the primaries of binaries can be different. In order to understand its influence on our results, we fixed our trained MLP model, but vary the IMF distribution in the test samples. We construct a test sample, in which the masses of the single stars and primaries of binaries follow a flat distribution. In Fig. 10, we compare the results of this case with the case with a flat distribution in logarithmic space. From the plot, we can find that the difference is very small. In other word, the IMF of single stars and the primaries of binaries has no influence on our results.

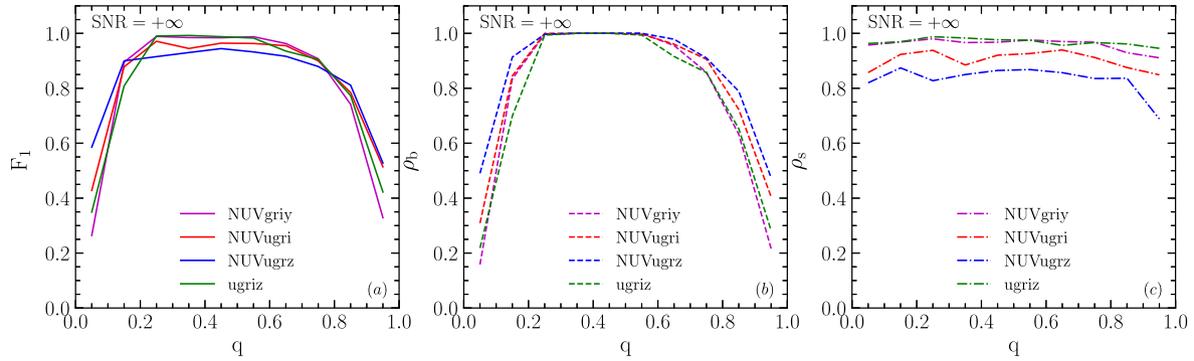

**Figure 11.** Dependence of $F_1$ (left-hand panel), $\rho_b$ (middle panel), and $\rho_s$ (right-hand panel) on the mass ratio $q$. Here we only show the case with SNR $= +\infty$ since other cases are very similar. Different colours are for samples with different combination of magnitudes.

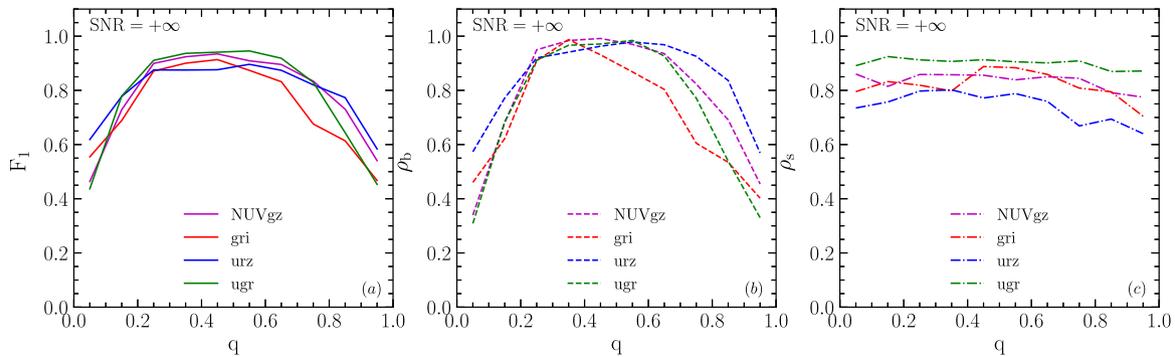

**Figure 12.** Similar to Fig. 11, but here the training and test sample only have 3 mag.

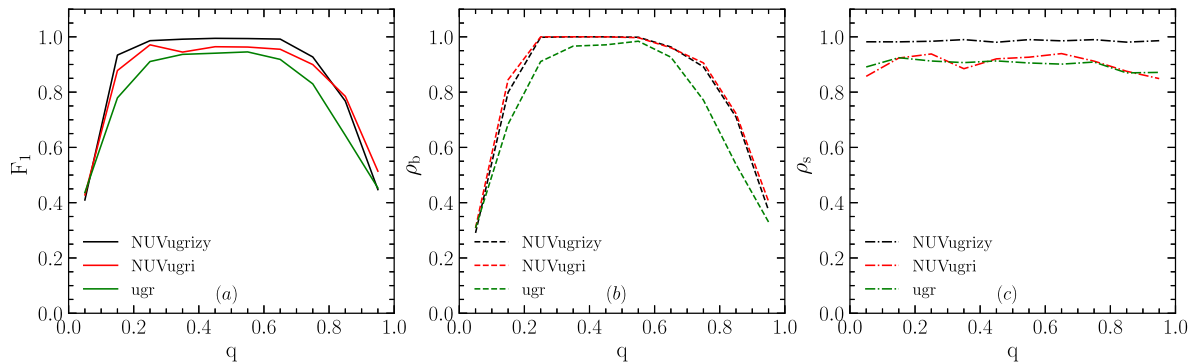

**Figure 13.** Comparison of results for training and test samples with different numbers of magnitudes. The black, red, and blue colours are for cases with 7, 5, and 3 mag. The left-hand, middle, and right-hand panels show the dependence of $F_1$, $\rho_b$, and $\rho_s$ on mass ratio, respectively.

### 5.4 Influence of the number of input magnitudes in the training and test samples

At the early stage of the CSST survey, the magnitudes at the seven bands may be not available simultaneously. Therefore, it is necessary to know whether our method still works if there are fewer input parameters. Following the method in Section 3.2, we construct several training and test samples with a sample number of magnitudes, i.e. 3 or 5 different magnitudes. With these samples, we train our MLP models and obtain the results about the performance of the trained MLP models. In Figs 11 and 12, we compare the results for different combinations of 5 or 3 mag. From these plots, we can find some differences between samples with a same number but different combinations of magnitudes. In Fig. 13, we compare the results for samples with different numbers of magnitudes, i.e. 7, 5, and 3 mag.

It is not out of expected that the MLP model performed better if there are more input magnitudes in the training and test samples. In addition, the MLP model still can have a reliable performance within the range of $0.3 < q < 0.7$ even the number of magnitudes is reduced to 3.

### 6 CONCLUSION AND OUTLOOK

In this work, we have developed a method to separate single MS stars from double MS binaries from the CSST survey with machine learning. Our method is reliable and efficient in identifying binaries with mass ratios between 0.20 and 0.80, which is independent of the initial mass ratio distribution. But the number of binaries identified with this method is not a good approximation to the number of binaries in the original sample. Therefore, we have improved this point by using an empirical mass ratio and detection efficiency (see equation 8). In addition, we find that our method cannot be applied to an observational sample with several systems smaller than 500.

When the CSST starts operating, we can use our method to identify candidates of double MS binaries and obtain binary fractions in the sample. Then we can study the dependence of binary fraction on stellar type, effective temperature and population age if these parameters can be obtained. Furthermore, we can study the distribution of binary parameters (e.g. primary mass, mass ratio, orbital period, and eccentricity) if we can infer these parameters by combining the data from the CSST with data from other telescopes, such as LAMOST and *Gaia*.

## ACKNOWLEDGEMENTS

We thank Yaotian Zeng for his help in data processing technology and Heran Xiong for the helpful discussion on data simulation. This work is supported by the National Natural Science Foundation of China (grants Nos 12125303, 12288102, and 12090040/3), the National Key R&D Program of China (grant No. 2021YFA1600403), the Yunnan Fundamental Research Projects (grants Nos. 202201BC070003), the International Centre of Supernovae, Yunnan Key Laboratory (No. 202302AN360001), and the Yunnan Revitalization Talent Support Programme Science & Technology Champion Project (grant NO. 202305AB350003). This work is also supported by the China Manned Space Project (No. CMS-CSST-2021-A10).

## DATA AVAILABILITY

The data underlying this paper will be shared on reasonable request to the corresponding author.

## REFERENCES

Abt H. A., Levy S. G., 1976, ApJS, 30, 273
Acernese F. et al., 2015, Class. Quant. Grav., 32, 024001
Allard F., Homeier D., Freytag B., Schaffenberger W., Rajpurohit A. S., 2013, Mem. Soc. Astron. Ital. Suppl., 24, 128
Allen P. R., 2007, ApJ, 668, 492
Amaro-Seoane P. et al., 2022, Living Rev. Relativ., 26, 2
Andrews J. J., Chanamé J., Agüeros M. A., 2017, MNRAS, 472, 675
Badenes C. et al., 2018, ApJ, 854, 147
Belczynski K., Kalogera V., Bulik T., 2002, ApJ, 572, 407
Bethe H. A., Brown G. E., 1998, ApJ, 506, 780
Bloom J. S., Sigurdsson S., Pols O. R., 1999, MNRAS, 305, 763
Cao Y. et al., 2018, MNRAS, 480, 2178
Chen X., Han Z., 2008, MNRAS, 387, 1416
Chinchor N., 1992, Proceedings of the 4th Conference on Message Understanding. Association for Computational Linguistics, McLean, Virginia, p. 22
Choi J., Dotter A., Conroy C., Cantiello M., Paxton B., Johnson B. D., 2016, ApJ, 823, 102
Clark B. M., Blake C. H., Knapp G. R., 2012, ApJ, 744, 119
Clem J. L., Landolt A. U., Hoard D. W., Wachter S., 2011, AJ, 141, 115
De Donder E., Vanbeveren D., 1998, A&A, 333, 557
Delfosse X. et al., 2004, in Hilditch R. W., Hensberge H., Pavlovski K., eds, ASP Conf. Ser. Vol. 318, Spectroscopically and Spatially Resolving the Components of the Close Binary Stars. Astron. Soc. Pac., San Francisco, p. 166
Dotter A., 2016, ApJS, 222, 8
Duchêne G., Kraus A., 2013, ARA&A, 51, 269
Duquennoy A., Mayor M., 1991, A&A, 248, 485
El-Badry K., Rix H.-W., 2019, MNRAS, 482, L139
El-Badry K. et al., 2018, MNRAS, 476, 528
Gao S., Liu C., Zhang X., Justham S., Deng L., Yang M., 2014, ApJ, 788, L37
Geller A. M., Latham D. W., Mathieu R. D., 2015, AJ, 150, 97
Gong Y. et al., 2019, ApJ, 883, 203
Guo Y. et al., 2022a, Res. Astron. Astrophys., 22, 025009
Guo Y., Liu C., Wang L., Wang J., Zhang B., Ji K., Han Z., Chen X., 2022b, A&A, 667, A44
Han Z., 1998, MNRAS, 296, 1019
Han Z., Podsiadlowski P., Maxted P. F. L., Marsh T. R., Ivanova N., 2002, MNRAS, 336, 449
Han Z., Podsiadlowski P., Maxted P. F. L., Marsh T. R., 2003, MNRAS, 341, 669
Han Z.-W., Ge H.-W., Chen X.-F., Chen H.-L., 2020, Res. Astron. Astrophys., 20, 161
Heintz W. D., 1969, J. R. Astron. Soc. Canada, 63, 275
Hurley J., Tout C. A., 1998, MNRAS, 300, 977
Hwang H.-C., Ting Y.-S., Zakamska N. L., 2022, MNRAS, 512, 3383
Kagra Collaboration, 2019, Nat. Astron., 3, 35
Li C., de Grijs R., Deng L., 2014, in de Grijs R., ed., Binary Systems, Their Evolution and Environments. p. 22, available at: https://academic.oup.com/mnras/article-pdf/436/2/1497/3932457/stt1669.pdf
Li L., Shao Z., Li Z.-Z., Yu J., Zhong J., Chen L., 2020, ApJ, 901, 49
Li J. et al., 2022a, MNRAS, 515, 3370
Li J. et al., 2022b, ApJ, 933, 119
Li J., Liu C., Zhang Z.-Y., Tian H., Fu X., Li J., Yan Z.-Q., 2023, Nature, 613, 460
LIGO Scientific Collaboration et al., 2015, Class. Quant. Grav., 32, 074001
Lipunov V. M., Postnov K. A., Prokhorov M. E., 1997, MNRAS, 288, 245
Lü G., Yungelson L., Han Z., 2006, MNRAS, 372, 1389
Mazzola C. N. et al., 2020, MNRAS, 499, 1607
Milone A. P. et al., 2012, A&A, 540, A16
Moe M., Di Stefano R., 2017, ApJS, 230, 15
Montgomery K. A., Marschall L. A., Janes K. A., 1993, AJ, 106, 181
Nelemans G., Yungelson L. R., Portegies Zwart S. F., Verbunt F., 2001, A&A, 365, 491
Niu H., Wang J., Fu J., 2020, ApJ, 903, 93
Pfahl E., Rappaport S., Podsiadlowski P., 2003, ApJ, 597, 1036
Portegies Zwart S. F., Yungelson L. R., 1998, A&A, 332, 173
Price-Whelan A. M. et al., 2020, ApJ, 895, 2
Raghavan D. et al., 2010, ApJS, 190, 1
Reino S., de Bruijne J., Zari E., d'Antona F., Ventura P., 2018, MNRAS, 477, 3197
Rumelhart D. E., Hinton G. E., Williams R. J., 1986, Nature, 323, 533
Sana H. et al., 2013, in Pugliese G., de Koter A., Wijburg M., eds, ASP Conf. Ser. Vol. 470, 370 Years of Astronomy in Utrecht. Astron. Soc. Pac., San Francisco, p. 141
Sollima A., Beccari G., Ferraro F. R., Fusi Pecci F., Sarajedini A., 2007, MNRAS, 380, 781
Sollima A., Carballo-Bello J. A., Beccari G., Ferraro F. R., Pecci F. F., Lanzoni B., 2010, MNRAS, 401, 577
Wang B., Han Z., 2012, New A Rev., 56, 122
Yuan H., Liu X., Xiang M., Huang Y., Chen B., Wu Y., Hou Y., Zhang Y., 2015, ApJ, 799, 135
Yungelson L., Livio M., 1998, ApJ, 497, 168
Yungelson L., Livio M., Tutukov A., Kenyon S. J., 1995, ApJ, 447, 656
Zhan H., 2011, Sci. Sin. Phys. Mech. Astron., 41, 1441

This paper has been typeset from a T<sub>E</sub>X/LAT<sub>E</sub>X file prepared by the author.